\pdfoutput=1
\documentclass[11pt]{article}

\usepackage[margin=1in]{geometry}
\usepackage{graphicx}
\usepackage{booktabs}
\usepackage{tabularx}
\usepackage{array}
\usepackage{enumitem}
\usepackage{float}
\usepackage{caption}
\usepackage{amsmath}
\usepackage{hyperref}
\usepackage{microtype}

\hypersetup{
    colorlinks=true,
    linkcolor=blue,
    citecolor=blue,
    urlcolor=blue,
    pdftitle={Zhinong AI: A Design-Science Study of an AI-Enabled Agricultural Decision-Support Platform for Smallholder Production},
    pdfauthor={Zhaoyang Li, Jiaqi Liu, Ruijie Zhang}
}

\newcolumntype{Y}{>{\raggedright\arraybackslash}X}
\setlist[itemize]{leftmargin=1.2em, itemsep=2pt, topsep=3pt}
\setlist[enumerate]{leftmargin=1.3em, itemsep=2pt, topsep=3pt}
\title{Zhinong AI: A Design-Science Study of an AI-Enabled Agricultural Decision-Support Platform for Smallholder Production}

\author{
Zhaoyang Li\\
University of Sanya\\
\and
Jiaqi Liu\\
Hebei International Studies University\\
\and
Ruijie Zhang\\
University of Sanya
}

\date{May 2026}

\begin{document}
\maketitle

\begin{abstract}
Artificial intelligence is increasingly moving from single-purpose agricultural recognition tools toward integrated decision-support systems that connect information access, diagnosis, task execution and post-action feedback. This paper presents a design-science case study of the Zhinong AI Agricultural Decision Platform, a farmer-facing system that integrates agricultural information push services, natural-language question answering, image-based crop disease diagnosis, plot and farming-calendar management, workflow orchestration, a Hainan Free Trade Port agricultural service zone and an age-friendly care mode. Based on public project materials, policy context and prior research on smart agriculture, machine learning and design science, the paper constructs a layered system architecture and a closed-loop decision process summarized as sensing, analysis, planning, execution and feedback. It further proposes a function-pain-point mapping matrix, an evaluation indicator system and a governance framework covering data provenance, model risk, expert review, privacy and adoption risk. The study does not claim measured field performance because production logs, controlled user studies and expert-labeled local image datasets were not available at the time of writing. Instead, the contribution is a structured research framework for transforming an AI agricultural prototype into an empirically testable, accountable and localized decision-support infrastructure for smallholder production.
\end{abstract}


\section{Introduction}
Agricultural production involves continuous decision-making under uncertainty. Farmers must respond to weather variability, market price changes, policy windows, crop growth stages, pest and disease symptoms, input costs and labor constraints. In many smallholder contexts, these decisions are still supported mainly by personal experience, fragmented online information and intermittent offline extension services. As a result, farmers may face delayed information access, unstable diagnostic judgment and weak follow-through after receiving advice.

The development of mobile computing, computer vision, natural-language processing and large language models creates a new opportunity for agricultural services. A practical agricultural AI platform should not merely provide isolated answers. It should help farmers transform scattered information into concrete actions, record execution, track outcomes and support later review. This paper therefore treats AI-enabled agriculture as a socio-technical decision-support problem rather than only as a model-accuracy problem.

The research object is the Zhinong AI Agricultural Decision Platform. Public project materials describe Zhinong AI as a farmer-oriented agricultural digital product that includes agricultural information push services, AI question answering, image-based disease diagnosis, plot and farming-calendar management, workflow mechanisms, a Hainan Free Trade Port agricultural service zone and an age-friendly care mode \cite{zhinong_home,sohu_release,sohu_team}. Public reports state that the platform is designed around common production pain points such as slow information acquisition, difficult disease identification and the lack of closed-loop production management \cite{sohu_release,sohu_team}.

This paper asks three research questions:
\begin{enumerate}
    \item How can the main information and decision pain points in grassroots agricultural production be mapped to platform functions?
    \item What system architecture and closed-loop decision process can organize Zhinong AI as an accountable agricultural decision-support artifact?
    \item Before large-scale field data are available, what evaluation indicators and governance mechanisms can guide subsequent empirical validation?
\end{enumerate}

The contribution is threefold. First, the paper translates a public platform case into a design-science framework that connects farmer pain points, platform modules, information flows and governance controls. Second, it proposes a closed-loop agricultural decision process that links information sensing, AI-assisted analysis, plan generation, task execution and outcome feedback. Third, it specifies an empirical validation roadmap for moving from design plausibility toward evidence-based claims about accuracy, usability, adoption and operational value.

\section{Background and Related Work}
\subsection{Digital agriculture and decision support}
Smart agriculture uses information and communication technologies, sensors, cloud platforms, data resources and AI to monitor, analyze and optimize agricultural production. Prior work on big data in smart farming emphasizes that data from sensors, machines and external services can strengthen decision-making, while also creating challenges around governance, ownership, privacy and business models \cite{wolfert2017big}. For smallholders, however, the core issue is often not the absence of advanced technology but the lack of a low-threshold service loop that converts timely information into understandable and executable actions.

China's digital-village policy context also reinforces the need for farmer-facing digital infrastructure. The Digital Village Development Action Plan (2022-2025) emphasizes digital infrastructure upgrading, smart-agriculture innovation and the broader digital transformation of rural production and governance \cite{xinhua2022digital}. In this context, an agricultural AI platform should be assessed not only by technical sophistication but also by accessibility, traceability and fit with local production systems.

\subsection{Machine learning in agriculture}
Machine learning has been applied to crop management, disease detection, weed detection, yield prediction, irrigation scheduling and soil management \cite{liakos2018ml}. Deep learning has been especially visible in plant disease identification. Mohanty et al. demonstrated the potential of convolutional neural networks for image-based plant disease detection under controlled image conditions \cite{mohanty2016plant}. Survey work also shows that deep learning can support agricultural recognition and prediction tasks across images, sensors and remote sensing data \cite{kamilaris2018deep}.

However, field deployment is harder than benchmark classification. Real field images vary in lighting, angle, occlusion, disease stage, crop variety and background clutter. Symptoms may also be caused by disease, pests, nutrient deficiency, drought stress or chemical injury. Therefore, an agricultural diagnosis module should present AI output as suspected diagnosis plus confidence, risk warning, alternative explanations and follow-up instructions rather than as an unquestioned final conclusion.

\subsection{Design science and human-centered agricultural AI}
Design science research studies the construction and evaluation of information-system artifacts intended to solve real-world problems. Hevner et al. argue that design science should produce useful artifacts and evaluate them rigorously in relation to the problem environment \cite{hevner2004design}. Peffers et al. further propose a design-science research methodology that moves from problem identification to objectives, design, demonstration, evaluation and communication \cite{peffers2007dsrm}. This paper adopts that logic: Zhinong AI is analyzed as a prototype-like artifact whose value depends on architecture, workflow, evaluation and governance.

Human-centered design is also essential. Farmer-facing AI systems must support users with different literacy levels, device familiarity and trust expectations. The care-mode design reported for Zhinong AI, including larger fonts, larger clickable areas, simplified entrances and higher contrast, reflects an accessibility requirement rather than a peripheral interface feature \cite{sohu_release,sohu_team}. In agricultural production, usability affects not only satisfaction but also whether advice is actually translated into timely field action.

\section{Research Object and Methodology}
\subsection{Research object}
The Zhinong AI Agricultural Decision Platform is positioned as a digital decision-support tool for grassroots agricultural production. According to public materials, the platform sets up an ``Agricultural Daily'' to push weather alerts, market information, policy subsidies and farming-calendar reminders based on region, crop and user interests. It also supports natural-language consultation and crop image upload for suspected pest or disease diagnosis. Additional modules include plot management, workflow center, a Hainan Free Trade Port agricultural service zone and an age-friendly care mode \cite{zhinong_home,sohu_release,sohu_team}.

The paper abstracts the platform into three application scenarios. The first is routine information service, including weather, prices, policy notices and farming reminders. The second is emergency diagnosis service, including suspected pest and disease recognition, severity description, recommended measures and follow-up review. The third is continuous production management, including plot records, input records, to-do tasks, workflow tracking and outcome review.

\subsection{Methodology and evidence boundary}
This study uses a design-science case-analysis method. The materials include the public project website, public media reports, digital-village policy information and prior literature on smart agriculture, AI and information-systems design. The analysis proceeds in four steps:
\begin{enumerate}
    \item extract major farmer pain points and platform functions from public materials;
    \item construct a function-pain-point mapping matrix;
    \item design a layered architecture and a closed-loop decision process;
    \item propose evaluation indicators, risk controls and an empirical validation roadmap.
\end{enumerate}

A clear evidence boundary is necessary. The study did not access backend logs, real user interviews, expert-labeled local image datasets or controlled field-trial results. Therefore, all numeric scores in the heatmap and radar figure are analytic design ratings used to demonstrate the evaluation framework. They should not be interpreted as measured user satisfaction, diagnostic accuracy or economic benefit.

\begin{figure}[H]
    \centering
    \includegraphics[width=0.98\linewidth]{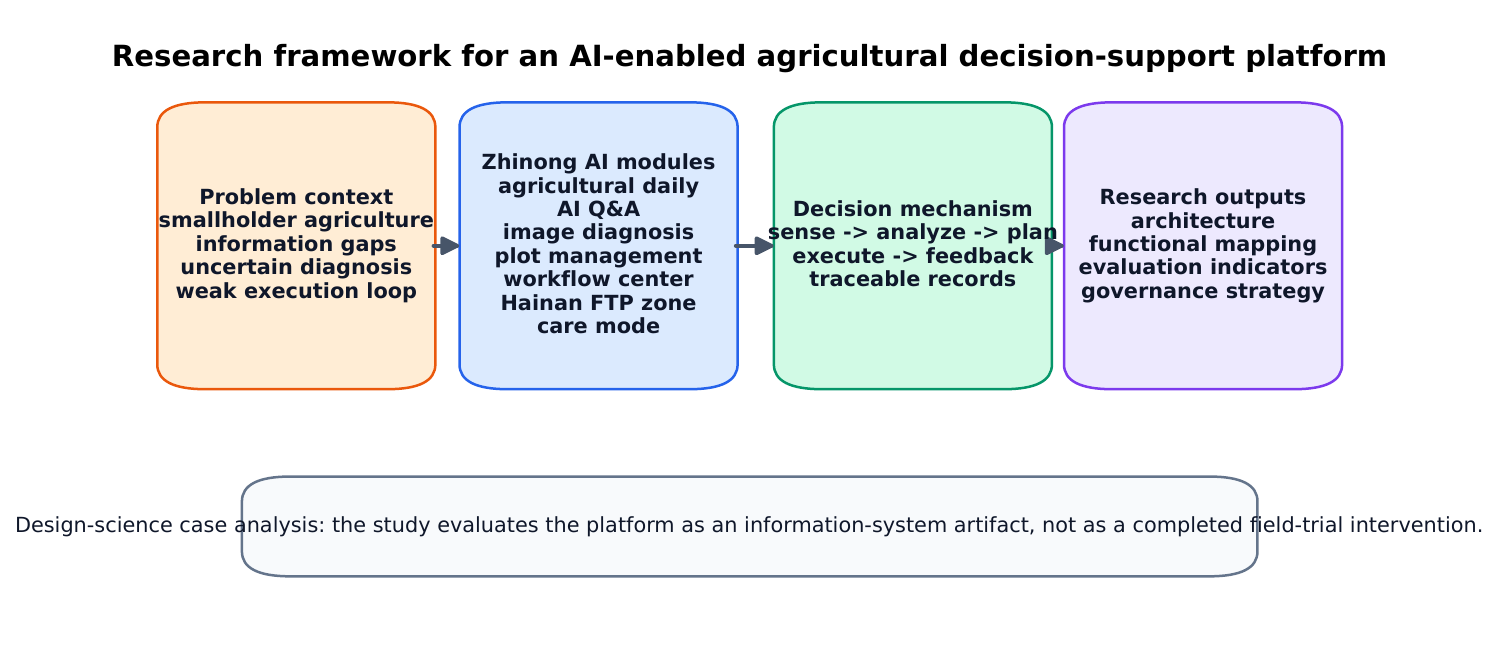}
    \caption{Research framework and platform overview. The diagram synthesizes public project information and the design-science analysis of this paper.}
    \label{fig:framework}
\end{figure}

Figure~\ref{fig:framework} summarizes the research framework.

\section{Requirements Analysis and System Architecture}
\subsection{Farmer decision pain points}
Grassroots agricultural decisions have strong contextual features. First, information is scattered and time-sensitive. Weather alerts, market price changes, subsidy windows and disease outbreaks all require rapid response. Second, agronomic knowledge is experiential and local. The same symptom may result from disease, insects, nutrient deficiency, water stress or chemical damage. Third, production execution often lacks process tools. A farmer may receive a suggestion but still fail to implement it because reminders, dosage cautions, input records and follow-up diagnosis are missing. Fourth, many rural users need more accessible interfaces, especially older farmers with limited digital-tool experience. Fifth, local policy and market rules are complex, particularly in regional settings such as Hainan tropical agriculture and the Free Trade Port.

\subsection{Function design}
Table~\ref{tab:functions} maps the main platform modules to target pain points and research significance. The modules should be read as a connected service system rather than as isolated features.

\begin{table}[H]
\centering
\caption{Zhinong AI modules, target pain points and research significance.}
\label{tab:functions}
\small
\renewcommand{\arraystretch}{1.18}
\begin{tabularx}{\linewidth}{p{0.20\linewidth}Y Y}
\toprule
\textbf{Module} & \textbf{Target pain point} & \textbf{Research significance} \\
\midrule
Agricultural Daily & Delayed access to weather, market, policy and farming-calendar information & Tests whether proactive information push can reduce search cost and improve risk awareness. \\
AI question answering & Fragmented agricultural knowledge and difficulty translating technical terms into actions & Connects agronomic knowledge retrieval with farmer-oriented explanation. \\
Image diagnosis & Difficulty identifying pests, diseases and abnormal crop symptoms in the field & Provides a computer-vision entry point for suspected diagnosis and follow-up record creation. \\
Plot management & Dispersed plot, input, cost and production-stage records & Converts advice into plot-level task management and cost accounting. \\
Workflow center & Complex tasks lack step-by-step execution and traceability & Structures planting plans, disease control and policy applications as auditable workflows. \\
Hainan FTP service zone & Local policies, tropical products, import-export rules and breeding information are difficult to track & Demonstrates regional knowledge-service adaptation for Hainan agriculture. \\
Care mode & Digital interfaces are difficult for older or low-literacy users & Embeds accessibility into agricultural AI adoption rather than treating it as a cosmetic feature. \\
\bottomrule
\end{tabularx}
\end{table}

\subsection{Layered architecture}
The platform can be organized into six layers: user layer, interaction layer, application layer, intelligent service layer, data resource layer and governance and security layer. The user layer includes farmers, cooperatives, extension workers and administrators. The interaction layer supports text, image and potentially voice input, plus alerts and care mode. The application layer hosts Agricultural Daily, Q\&A, Image Diagnosis, Plot Management, Workflow Center and the Hainan service zone. The intelligent service layer combines retrieval-augmented generation, visual recognition, rule-based checks, recommendation and confidence reporting. The data resource layer connects weather, market, policy, agronomic knowledge, plot records and diagnostic images. The governance layer manages source labels, audit logs, privacy permissioning, expert review and risk disclaimers.

\begin{figure}[H]
    \centering
    \includegraphics[width=0.96\linewidth]{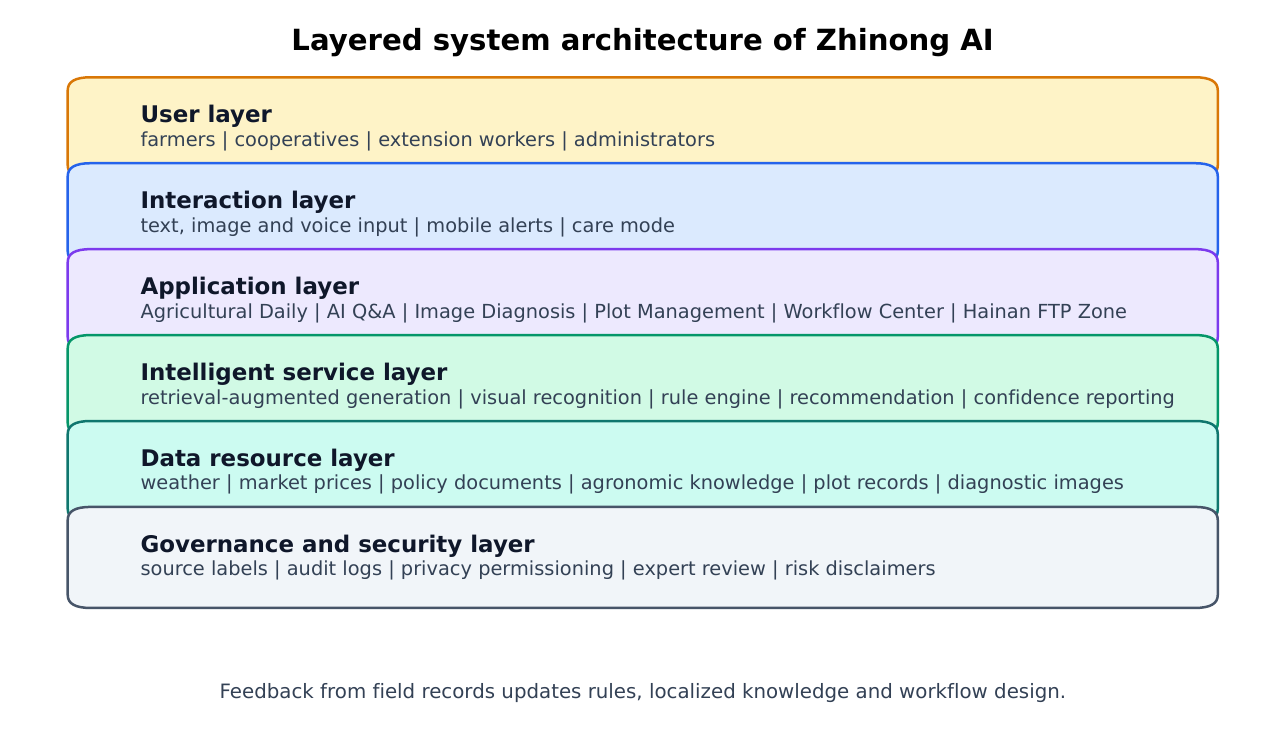}
    \caption{Layered system architecture of Zhinong AI. The structure emphasizes the integration of application modules, intelligent services, data resources and governance controls.}
    \label{fig:architecture}
\end{figure}

\section{Technical Route and Implementation Mechanisms}
\subsection{Multi-source information push and provenance labels}
The Agricultural Daily can serve as the platform's active-service entrance. It should aggregate weather alerts, market prices, policy subsidies and farming reminders based on location, crop type, plot cycle and user preference. To avoid information overload, the ranking strategy should prioritize items by urgency and relevance. Each message should explain why it matters for a specific crop or plot and provide source, release time and applicable region. Provenance labels are important because farmers need to verify high-impact advice, especially when the information affects pesticide use, harvesting, sales or subsidy application.

\subsection{Agricultural question answering with retrieval and rule checks}
The Q\&A module should not rely only on a general-purpose language model. A safer route is retrieval-augmented generation: retrieve relevant passages from agricultural manuals, policy documents, disease databases and local extension materials, generate a concise answer, and then apply rule checks and risk prompts. High-risk topics such as pesticide dosage, restricted chemicals, harvest safety intervals and subsidy eligibility require stricter constraints or human review.

\subsection{Image-based diagnosis and follow-up records}
The image-diagnosis pipeline can include image-quality checks, crop-part recognition, lesion feature extraction, disease classification, confidence output, advice generation and follow-up reminders. Since field images are noisy, the platform should avoid presenting AI output as a definitive diagnosis. A better output format is suspected diagnosis, confidence level, visually similar alternatives, immediate field action suggestions, and a prompt to consult an extension worker when risk is high. Each diagnosis should form a record containing image, date, plot, crop stage, suggested measure and follow-up result. Such records support later model evaluation and accountability.

\subsection{Plot management and workflow orchestration}
Plot management converts AI advice into executable production tasks. After farmers enter plot area, crop variety, sowing date, soil condition, fertilizer and pesticide application, labor input and cost information, the platform can generate to-do items according to crop stage and local farming calendar. The workflow center further decomposes complex matters into information collection, plan generation, review, execution reminder and outcome tracking. This mechanism gives planting plans, disease control, policy applications and research-material analysis continuity and traceability.

\subsection{Regional service and age-friendly design}
The Hainan Free Trade Port service zone reflects a regional knowledge-service strategy. Tropical crops, breeding activities, import-export compliance and local policy updates require localized databases and frequent updating. The care mode supports digital inclusion by using larger type, higher contrast, fewer levels, clearer prompts and simplified entrances. The usability of such a mode should be evaluated by task completion rate, mis-tap rate, reading comprehension and willingness to reuse, not merely by interface appearance.

\section{Closed-loop Decision Process}
The key question for agricultural AI is not simply whether the system can answer a question, but whether it can support a cycle from information to action and then to learning. This paper summarizes the Zhinong AI mechanism as five stages: sense, analyze, plan, execute and feedback.

In the sensing stage, the platform collects weather, market, policy, plot and image information. In the analysis stage, it applies knowledge retrieval, vision models and rule checks. In the planning stage, it produces suspected diagnoses, recommendations, warnings or task lists. In the execution stage, to-do reminders and workflow forms support implementation. In the feedback stage, follow-up diagnosis, result records and audit trails inform later recommendations.

\begin{figure}[H]
    \centering
    \includegraphics[width=0.98\linewidth]{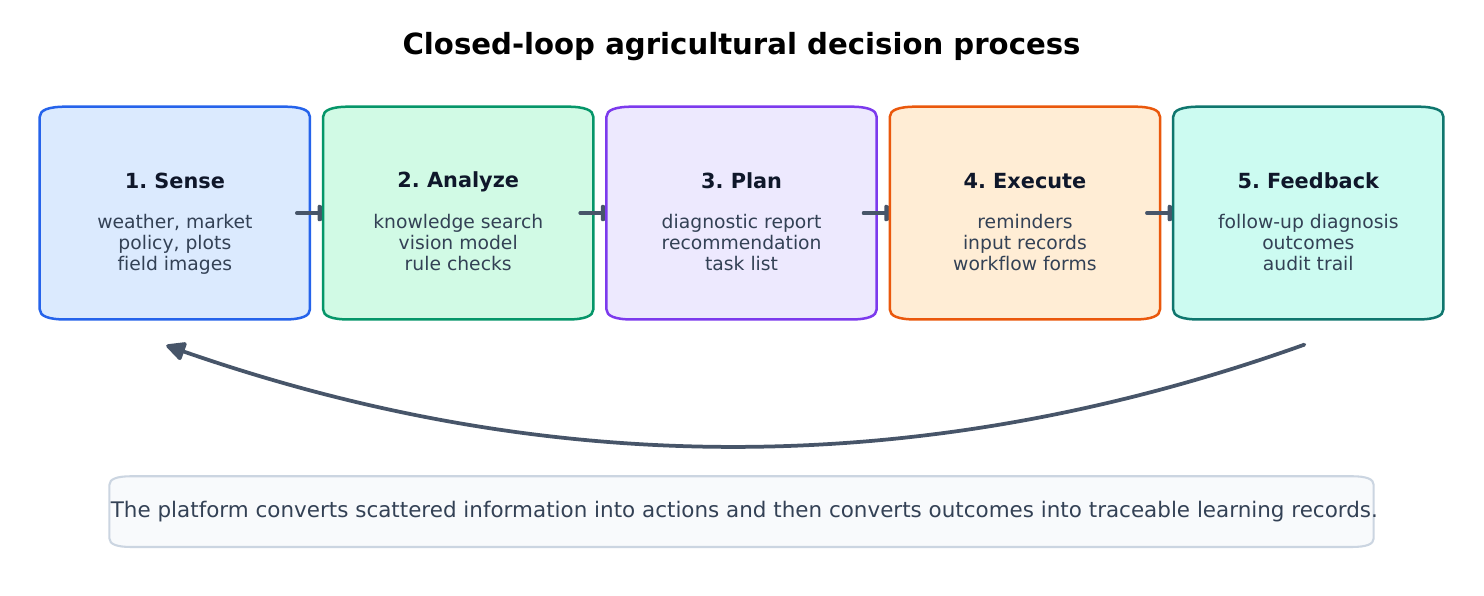}
    \caption{Closed-loop agricultural decision process. The intended value is a traceable cycle rather than a one-time answer.}
    \label{fig:loop}
\end{figure}

This loop is important because agricultural value is created in execution. If diagnosis results remain disconnected from plot records and follow-up tasks, the platform may become only a consultation tool. If diagnosis, task execution and outcome feedback are linked, the platform can gradually accumulate local knowledge and improve trust.

\section{Function-Pain-Point Mapping and Evaluation Framework}
\subsection{Analytic support matrix}
\begin{figure}[H]
    \centering
    \includegraphics[width=0.90\linewidth]{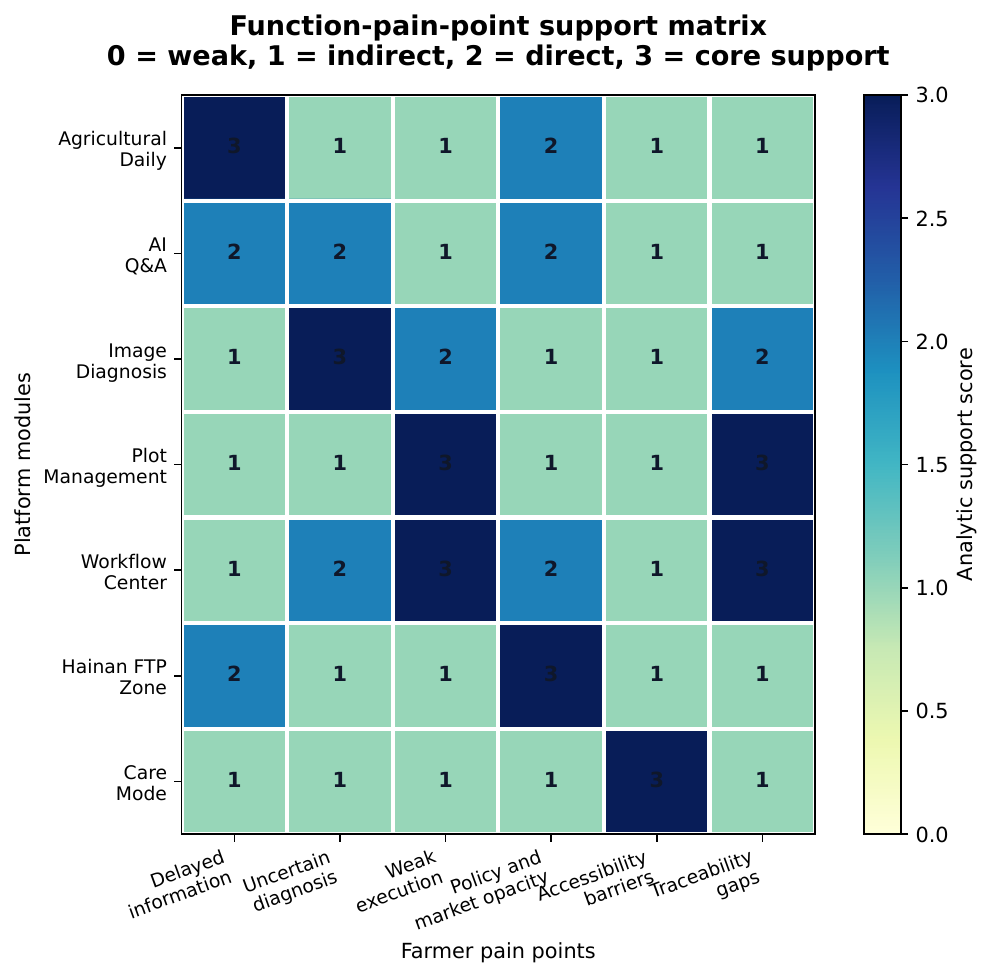}
    \caption{Function-pain-point support matrix. The numeric values are analytic design scores, not empirical measurements.}
    \label{fig:matrix}
\end{figure}

Figure~\ref{fig:matrix} presents a function-pain-point support matrix. A score of 0 indicates weak relevance, 1 indicates indirect support, 2 indicates direct support and 3 indicates core support. The scores are based on public platform descriptions and researcher judgment, not on measured user data.

The matrix indicates that Agricultural Daily is central to delayed information, image diagnosis is central to uncertain diagnosis, plot management and workflow center are central to execution and traceability, the Hainan service zone supports policy and market opacity, and care mode directly addresses accessibility barriers. A major improvement direction is cross-module data linkage. For example, a weather warning from Agricultural Daily should automatically generate plot-level tasks, while a disease-diagnosis result should automatically enter a follow-up workflow.

\subsection{Evaluation indicators}
Table~\ref{tab:evaluation} proposes an evaluation indicator system for future empirical research. The indicators cover technical performance, user experience, process execution, adoption and governance.

\begin{table}[H]
\centering
\caption{Proposed empirical evaluation indicators for Zhinong AI.}
\label{tab:evaluation}
\small
\renewcommand{\arraystretch}{1.18}
\begin{tabularx}{\linewidth}{p{0.22\linewidth}Y Y}
\toprule
\textbf{Dimension} & \textbf{Possible indicators} & \textbf{Data source} \\
\midrule
Information reach & reading rate, click-through rate, alert response time, source-verification rate & platform logs, farmer surveys \\
Diagnostic support & top-1/top-3 accuracy, confidence calibration, false positive and false negative cases, expert review agreement & expert-labeled image dataset, audit sample \\
Task execution & to-do completion rate, overdue rate, input-record completeness, follow-up diagnosis rate & workflow logs, plot records \\
Usability & first-task completion rate, mis-tap rate, average task time, care-mode satisfaction & controlled usability test, A/B experiment \\
Adoption and trust & weekly active users, repeat consultation rate, perceived usefulness, willingness to recommend & surveys, interviews, logs \\
Economic and operational value & information-search time saved, diagnosis response time, input-use record quality, cost-accounting completeness & field pilot data, farmer diaries \\
Governance & percentage of answers with sources, high-risk answer review rate, privacy-permission compliance, deletion-request completion & audit logs, governance checklist \\
\bottomrule
\end{tabularx}
\end{table}

\subsection{Design-fit radar}
\begin{figure}[H]
    \centering
    \includegraphics[width=0.62\linewidth]{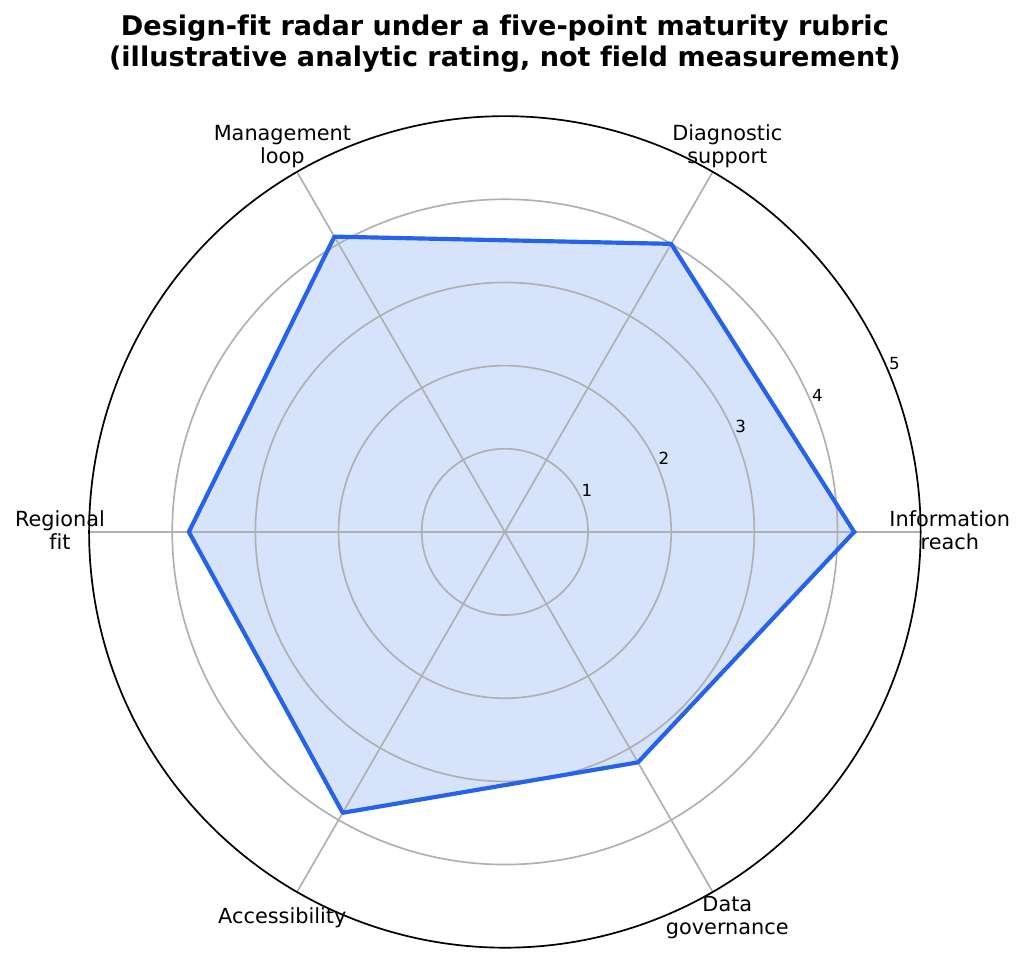}
    \caption{Design-fit radar under a five-point maturity rubric. The values are illustrative analytic ratings, not field measurements.}
    \label{fig:radar}
\end{figure}

Figure~\ref{fig:radar} gives an illustrative design-fit radar using a five-point maturity rubric. Information reach, diagnostic support and management loop are rated relatively high because they correspond to the platform's central modules. Data governance is rated lower to highlight the need for source verification, privacy protection, model evaluation, expert review and user authorization.

\section{Risks and Governance}
The deployment of agricultural AI creates value only if risks are managed. Table~\ref{tab:risks} summarizes key risks and governance strategies.

\begin{table}[H]
\centering
\caption{Platform deployment risks and governance strategies.}
\label{tab:risks}
\small
\renewcommand{\arraystretch}{1.18}
\begin{tabularx}{\linewidth}{p{0.20\linewidth}Y Y}
\toprule
\textbf{Risk type} & \textbf{Typical manifestation} & \textbf{Governance strategy} \\
\midrule
Data risk & expired policy information, unstable market data, biased image samples, incomplete plot records & source labels, update timestamps, data-quality checks, missing-data prompts \\
Model risk & misdiagnosis, hallucinated answer, poor confidence expression, weak cross-region transfer & confidence thresholds, expert review, fallback rules, post-deployment monitoring \\
Responsibility risk & AI advice affects pesticide use, loss claims or subsidy applications & risk-level classification, human confirmation, clear disclaimers, audit logs \\
Privacy risk & plot data, images and production records may reveal sensitive household or business information & minimal collection, user permissioning, anonymization, deletion mechanisms \\
Adoption risk & farmers may distrust AI, abandon the platform or use advice incorrectly & care mode, training, extension-worker mediation, feedback channels \\
Regional risk & general knowledge may ignore Hainan policy, tropical crops or compliance rules & local knowledge base, policy update workflow, regional expert collaboration \\
\bottomrule
\end{tabularx}
\end{table}

A practical governance chain should include data-source labeling, risk-level prompts, human review for high-risk advice, user feedback channels and log-based traceability. For pesticide use, disease outbreak warnings and policy applications, the system should either cite authoritative sources or trigger expert confirmation. For privacy, the platform should follow minimal necessary collection, tiered authorization, anonymized display and user-controlled deletion.

\section{Discussion}
The Zhinong AI case illustrates that the competitiveness of an agricultural AI platform does not come only from model capacity. It comes from coordination among models, data, workflows, regional knowledge and user experience. For farmers, useful AI output is concrete: whether to irrigate, whether to spray, when to apply for a subsidy, what to do next and how to record the result. A system limited to question answering or image recognition can be replaced by one-time consultation. A system that builds plot records, workflows and follow-up diagnosis can become part of daily production management.

The analysis also shows that agricultural AI should not overpromise automated decision-making. Agricultural environments vary significantly, responsibility chains are long and local expertise is important. AI is better positioned as an assistant for judgment, workflow management and record keeping. The design principles should be explainability, reviewability, traceability and correctability.

From a regional-service perspective, the Hainan Free Trade Port zone provides a path for differentiated development. Agricultural AI platforms that rely only on general knowledge may miss local policy, crop structure and market rules. By constructing regional knowledge bases and collaborating with local extension experts, a general agricultural assistant can evolve into local agricultural-service infrastructure.

\section{Empirical Validation Roadmap}
\begin{figure}[H]
    \centering
    \includegraphics[width=0.98\linewidth]{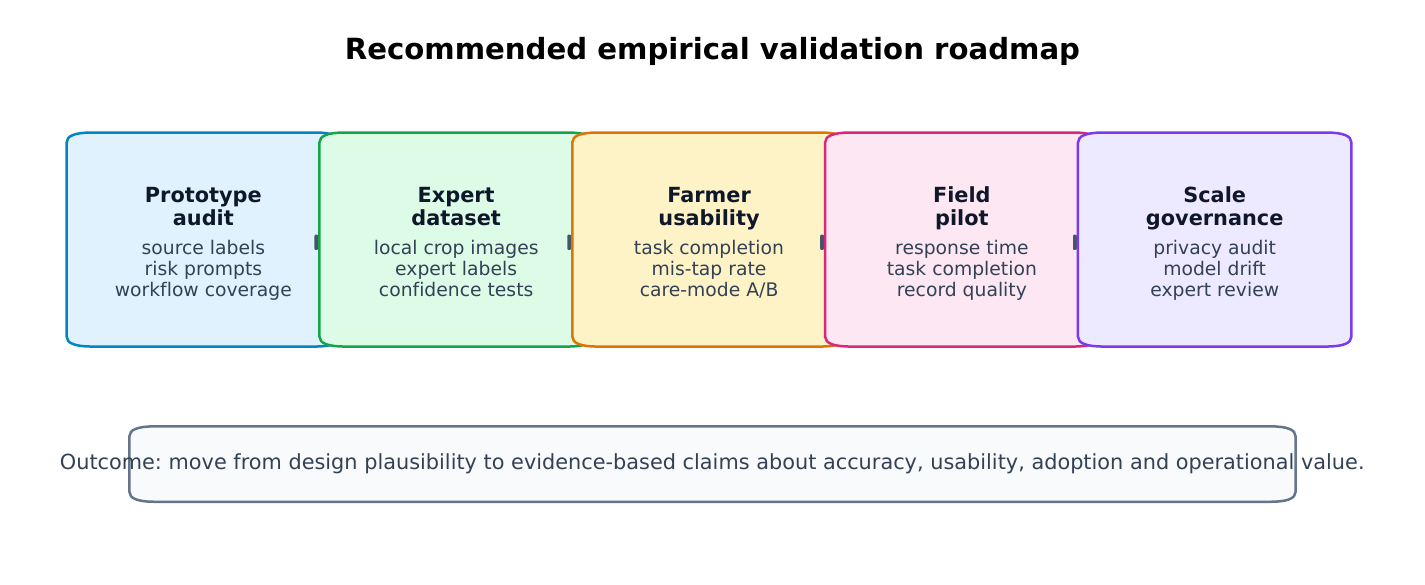}
    \caption{Recommended empirical validation roadmap for future work.}
    \label{fig:roadmap}
\end{figure}

To convert the present design analysis into a stronger empirical study, the next step is to conduct staged validation. Figure~\ref{fig:roadmap} recommends a roadmap from prototype audit to scale governance.

A feasible validation plan would include: (1) 30 to 100 farmer interviews and questionnaires measuring ease of use, trust and willingness to reuse; (2) an expert-labeled local crop disease dataset for measuring top-1/top-3 accuracy, calibration and error types; (3) continuous logs from the plot-management module for measuring task completion and input-record quality; (4) Agricultural Daily log metrics such as reading rate, alert-response time and source-verification behavior; and (5) a care-mode A/B test comparing task completion time, mis-tap rate and satisfaction between standard and age-friendly interfaces.

\section{Conclusion}
This paper analyzed the Zhinong AI Agricultural Decision Platform as a design-science case of AI-enabled agricultural decision support. The platform addresses delayed information, uncertain diagnosis, weak management closure, policy and market opacity and accessibility barriers through Agricultural Daily, AI question answering, image diagnosis, plot management, workflow center, Hainan Free Trade Port service zone and care mode.

The central framework is a closed-loop process of sense, analyze, plan, execute and feedback. The paper contributes a layered architecture, a function-pain-point matrix, an evaluation indicator system and a governance strategy. The theoretical value is to shift AI agriculture analysis from single-model evaluation toward an integrated framework of functional response, workflow closure and accountable governance. The practical value is to provide a roadmap for transforming Zhinong AI from a prototype-oriented service into an empirically validated and regionally adaptable agricultural decision-support infrastructure.

Future research should focus on three tasks: building local field-image datasets with expert labels, testing usability among farmers including older users, and constructing sustainable regional knowledge bases for tropical crops, policy, market and breeding services in Hainan and other local production systems.

\section*{Author Contributions}
Zhaoyang Li, Jiaqi Liu and Ruijie Zhang contributed to the conceptualization, platform case analysis and manuscript preparation. All authors reviewed the final preprint version.

\section*{Conflict of Interest}
The authors declare no competing interests in the preparation of this preprint.

\section*{Data Availability}
This manuscript is based on public project information, public reports and published literature. No human-subject dataset, platform backend log or private field dataset is included in this version.

\end{document}